\documentclass[%
 reprint,
 amsmath,amssymb,
 superscriptaddress,
 aps,
]{revtex4-2}

\usepackage{graphicx}
\usepackage{blindtext}
\usepackage{dcolumn}
\usepackage{bm}
\usepackage{braket}
\usepackage{setspace}
\usepackage{color}
\usepackage{mathtools}

\usepackage[english]{babel}

\begin{document}
\preprint{APS/123-QED}

\title{Can Constrained Quantum Annealing Be Effective in Noisy Quantum Annealers? }

\author{Ryoya Igata}
\affiliation{
Department of Computer Science and Engineering, Shibaura Institute of Technology, Toyosu, Tokyo 135-8548, Japan}

\author{Myonsok I}
\affiliation{
Department of Computer Science and Engineering, Shibaura Institute of Technology, Toyosu, Tokyo 135-8548, Japan}

\author{Yuya Seki}
\affiliation{Graduate School of Science and Technology, Keio University,
Kanagawa 223-8522, Japan}

\author{Yuta Mizuno} %
\affiliation{Research Institute for Electronic Science, Hokkaido University, Sapporo, Hokkaido 001-0020, Japan}
\affiliation{Institute for Chemical Reaction Design and Discovery (WPI-ICReDD), Hokkaido University, Sapporo, Hokkaido 001-0021, Japan}
\affiliation{Graduate School of Chemical Sciences and Engineering, Hokkaido University, Sapporo, Hokkaido 060-8628, Japan}

\author{Shohei Watabe}
\affiliation{
Department of Computer Science and Engineering, Shibaura Institute of Technology, Toyosu, Tokyo 135-8548, Japan}

\begin{abstract}
We investigate the performance of penalty-based quantum annealing (PQA) and constrained quantum annealing (CQA) in solving the graph partitioning problem under various noise models, including depolarizing, bit-flip, and phase-flip noise. 
We found that even in the absence of noise, 
the relative superiority of PQA or CQA is highly problem-dependent. 
PQA generally demonstrates relatively stable success probabilities, while CQA's performance varies significantly across problem instances. 
Notably, CQA experiences leakage from the constraint-satisfying subspace under most noise models, with the exception of phase-flip noise, where no leakage occurs exactly. The dependence of success probability on noise strength reveals that PQA consistently outperforms CQA in the weak noise regime, whereas CQA achieves higher success probabilities under strong phase-flip noise. 
However, even in this case, the overall success probability remains low, suggesting that CQA is not yet practical for noisy quantum annealers without the development of error-corrected quantum annealing and autonomous error correction techniques to address leakage from the constraint-satisfying subspace. 
To confine the system within the constraint subspace in CQA, error correction methods and cat code encoding will likely be required, both of which can suppress bit-flip errors. 
\end{abstract}


\maketitle


\section{Introduction}

Quantum annealing (QA)~\cite{kadowaki1998quantum} has garnered significant attention as a modern and practical approach for addressing combinatorial optimization problems (COPs)~\cite{lucas2014ising}. In particular, for COPs with constraint conditions, the use of a penalty Hamiltonian added to the problem Hamiltonian in QA is a conventional technique to ensure that the obtained solution satisfies the imposed constraints. This method effectively maps the constrained problem into an unconstrained one by penalizing violations of the constraints, drawing upon classical optimization strategies.

However, the use of a penalty Hamiltonian in QA is associated with several notable drawbacks~\cite{PhysRevApplied.5.034007}. First, its implementation is often hardware-inefficient, as the penalty Hamiltonian typically requires an all-to-all connectivity due to the fully connected nature of its graph representation. Realizing such a fully connected graph in logical qubits necessitates significant resource overhead for physical qubits. 

Second, the introduction of a penalty Hamiltonian imposes an additional artificial energy scale on the system. The two-body interactions required for all-to-all connectivity lead to a quadratic increase in the energy scale, which can surpass the energy scale of the original problem Hamiltonian~\cite{PhysRevApplied.5.034007}. Moreover, in practical applications, the optimal strength of the penalty may scale linearly with the degrees of freedom~\cite{PhysRevResearch.6.013115}, exacerbating the dominance of the penalty Hamiltonian's energy scale over that of the main problem Hamiltonian for large-size problems. In actual quantum devices, energy-scale normalization is required, and the precision of Hamiltonian parameter setting is finite; thus, such a significant imbalance between the energy scales of the main problem and penalty Hamiltonian may result in poor performance of the quantum annealing process.

To address these limitations, an alternative quantum approach known as constrained quantum annealing (CQA) has been proposed~\cite{PhysRevApplied.5.034007}. This method has been successfully applied to various problems, including graph partitioning~\cite{PhysRevApplied.5.034007} and graph coloring~\cite{PhysRevApplied.5.034007, PhysRevA.98.022301, doi:10.7566/JPSJ.89.064001}. In CQA, the initial state is prepared to satisfy the constraints, and the driver Hamiltonian is designed to commute with the operator corresponding to the constraints. Consequently, the quantum state remains within the constraint-satisfying subspace throughout the annealing process. Since CQA avoids the need for a penalty Hamiltonian, it circumvents the issues of hardware inefficiency and artificial energy scaling. Additionally, CQA operates within a reduced Hilbert space, making it more efficient for simulations and opening a larger energy gap in the searching Hilbert subspace compared to  the penalty-based quantum annealing (PQA), which explores the full Hilbert space~\cite{PhysRevA.98.022301}.

While these theoretical advantages are well-founded, their practical utility remains uncertain, as real-world quantum devices are inevitably subject to environmental noise, which could compromise their performance in experimental settings.  
This raises the important question of whether CQA can maintain its efficacy in the presence of noise. In particular, while CQA imposes constraints purely through quantum mechanical way without a penalty term, noise could cause the quantum state to drift out of the constraint-satisfying subspace. Although PQA approach is hardware-inefficient and suffers from energy scaling issues, it remains unclear how CQA performs in noisy quantum annealers.

In this paper, we analyze and compare the performance of PQA and CQA in solving the graph partitioning problem under various noise models, including depolarizing, bit-flip, and phase-flip noise. We examine how noise affects the success probability and leakage from the constraint-satisfying subspace in each approach. 
Our results show that, contrary to our expectations, 
CQA does not always outperform PQA even in the absence of noise. 
PQA maintains relatively stable performance, whereas CQA exhibits significant variability. CQA suffers from leakage out of the constraint-satisfying subspace, particularly under depolarizing and bit-flip noise. 
Interestingly, phase-flip noise does not induce leakage in CQA, and thus CQA demonstrates competitive success rates even under strong phase-flip noise.
However, this does not immediately imply that CQA is inherently superior, as the absolute success probabilities for both methods remain low under these conditions. In CQA, the development of error-corrected quantum annealing and autonomous error correction will be crucial to mitigate leakage from the constraint-satisfying subspace. In particular, the cat code is promising for CQA as it exponentially suppresses bit-flip errors with the cat size~\cite{Lescanne2020}.

\section{Formalism}

In this work, we focus on the graph partitioning problem for an undirected graph $G = (V, E)$, which is known to be NP-hard~\cite{inproceedings}.
Here, $V$ and $E$ denote, respectively, the vertex set and the edge set.
The objective of the problem is to partition the vertex set $V$ into two subsets such that the difference in the number of vertices between the two subsets is equal to a specified value, while minimizing the number of edges connecting vertices from different subsets.
The problem Hamiltonian can be formulated as~\cite{PhysRevApplied.5.034007}:
\begin{align}
\hat{H}_{\rm prob} = \sum_{(i,j) \in E} \frac{1}{2} \left( 1 - \hat{\sigma}_i^z \hat{\sigma}_j^z \right),
\end{align}
where $\hat{\sigma}_i^z$ represents the Pauli-$z$ operator for a vertex $i$. 
The constraint $ C(\{\hat{\sigma}_i^z\})$ for ensuring desired partitioning is expressed as:
\begin{align}\label{eq:partitioning_constraint}
 C(\{\hat{\sigma}_i^z\}) | \phi \rangle 
\coloneqq 
\biggl ( \sum_{i=1}^{n}  \hat{\sigma}_i^z \biggr ) | \phi \rangle = c | \phi \rangle,
\end{align}
where $c$ is an eigenvalue representing the required partition balance and $|\phi \rangle$ is its eigenstate. 
Here, we set $c=0$ throughout this paper, where we assume the equal balance case. 

In PQA, the total Hamiltonian is given by $\hat{H}_{\rm p} = \hat{H}_{\rm prob} + \alpha \hat{H}_{\rm pena}$, where $\hat{H}_{\rm pena}$ is the penalty Hamiltonian that enforces the constraint in Eq.~\eqref{eq:partitioning_constraint}:
\begin{align}
    \hat{H}_{\rm pena} = \left[C(\{\hat{\sigma}_i^z\}) - c\right]^2,
\end{align}
Here, $\alpha$ is penalty strength. 
The total time-dependent Hamiltonian during the annealing process is:
\begin{align}
    \hat{H}_{\rm QA}(t) = \frac{t}{T} \hat{H}_{\rm p} + \left( 1 - \frac{t}{T} \right) \hat{H}_{\rm d},
\end{align}
where $\hat{H}_{\rm d} = \sum_{i=1}^{n} \hat{\sigma}_i^x$ is the driver Hamiltonian, and $T$ is the annealing time. The penalty Hamiltonian $\hat{H}_{\rm pena}$ requires all-to-all connectivity, making its implementation challenging on current quantum devices~\cite{PhysRevApplied.5.034007}.

In contrast, CQA circumvents the need for a penalty Hamiltonian. In CQA, the total Hamiltonian is given by:
\begin{align}
    \hat{H}_{\rm CQA}(t) = \frac{t}{T} \hat{H}_{\rm prob} + \left( 1 - \frac{t}{T} \right) \hat{H}_{\rm d}^{\rm CQA},
\end{align}
where the driver Hamiltonian is defined as:
\begin{align}
    \hat{H}_{\rm d}^{\rm CQA} = - \sum_{i} \left( \hat{\sigma}_i^x \hat{\sigma}_{i+1}^x + \hat{\sigma}_i^y \hat{\sigma}_{i+1}^y \right).
\end{align}
This driver Hamiltonian commutes with the constraint operator, i.e., $[\hat{H}_{\rm d}^{\rm CQA}, C(\{\hat{\sigma}_i^z\})] = 0$. If the initial state is an eigenstate of $C(\{\hat{\sigma}_i^z\})$ that satisfies the constraint, the state remains within the constraint-satisfying subspace during the entire annealing process, as $[\hat{H}_{\rm CQA}(t), C(\{\hat{\sigma}_i^z\})] = 0$ at any $t$.

It should be noted that the ground state $|g\rangle$ of $\hat{H}_{\rm d}^{\rm CQA}$ does not necessarily satisfy the desired partitioning constraint $C(\{\hat{\sigma}_i^z\}) |g\rangle = c |g\rangle$. In this study, to ensure the initial state satisfies the constraint, we initialize the system using the ground state of the driver Hamiltonian with an added penalty term, $\hat{H}_{\rm d}^{\rm CQA} + \alpha_{\rm ini} \hat{H}_{\rm pena}$. This ensures that $C(\{\hat{\sigma}_i^z\}) |g\rangle = c |g\rangle$ holds at the start of the annealing process. (In practical implementation on quantum devices, more efficient methods for the initial state preparation should be considered, but it is beyond the scope of the present study.)

In PQA, the search for the ground state occurs within the full Hilbert space $\mathcal{H}$ spanned by the eigenstates of $\hat{H}_{\rm prob}$. In CQA, however, the annealing process is restricted to a subspace $\mathcal{H}_{\rm c}$ of the full Hilbert space, where $\mathcal{H}_{\rm c}$ consists of eigenstates $|i_{\rm c}\rangle$ that satisfy the constraint $C(\{\hat{\sigma}_i^z\}) |i_{\rm c}\rangle = c |i_{\rm c}\rangle$. This reduction in search space leads to increased simulation efficiency.

To study the impact of noise, we solve the Lindblad master equation:
\begin{align}
    \frac{\partial \rho}{\partial t} = -i [\hat{H}, \rho] + \mathcal{D}(\rho),
\end{align}
where $\rho$ is the density matrix, and $\mathcal{D}$ represents the incoherent processes described by the Lindblad superoperator:
\begin{align}\label{eq:lindbladian}
    \mathcal{D}(\rho) = \sum_k 
    \left ( 
    \hat{L}_k \rho \hat{L}_k^\dagger - \frac{1}{2} \left\{ \hat{L}_k^\dagger \hat{L}_k, \rho \right\}
    \right ),
\end{align}
with $\hat{L}_k$ being the Lindblad jump operators. Here, the curly bracket in Eq.~\eqref{eq:lindbladian} denotes anticommutator. We consider three noise models~\cite{PhysRevA.71.032350}: phase-flip noise ($\hat{L}_i = \sqrt{\gamma} \hat{\sigma}_i^z$), bit-flip noise ($\hat{L}_i = \sqrt{\gamma} \hat{\sigma}_i^x$), and depolarizing noise, where the jump operators are 
\begin{align}
    %
    \hat{L}_{(i,1)} = \frac{\sqrt{\gamma}}{2} \hat{\sigma}_i^x, 
    \,
    \hat{L}_{(i,2)} = \frac{\sqrt{\gamma}}{2} \hat{\sigma}_i^y, 
    \,
    \hat{L}_{(i,3)} = \frac{\sqrt{\gamma}}{2} \hat{\sigma}_i^z.
\end{align} 
In the case of depolarizing errors, the summation index $k$ in the Lindblad's master equation runs over 
$(i,\mu)$, where $\mu = 1,2,3$. 
The parameter $\gamma$ represents the noise strength, and the different noise models allow us to explore the robustness of CQA in various noisy environments.

\section{Results}

We evaluate the performance of PQA and CQA in addressing an 8-node graph partitioning problem under distinct noise models. Both approaches are executed with an annealing time of $T=20$. For the PQA, the penalty strength is fixed at $\alpha = 8$, whereas in the CQA, the initial penalty strength is set to $\alpha_{\rm ini} = 100$. A total of 100 random instances of the 8-node graph partitioning problem were generated for this study in performance comparison.


\begin{figure}[tbp]
    \centering
    \includegraphics[keepaspectratio, width=80mm]{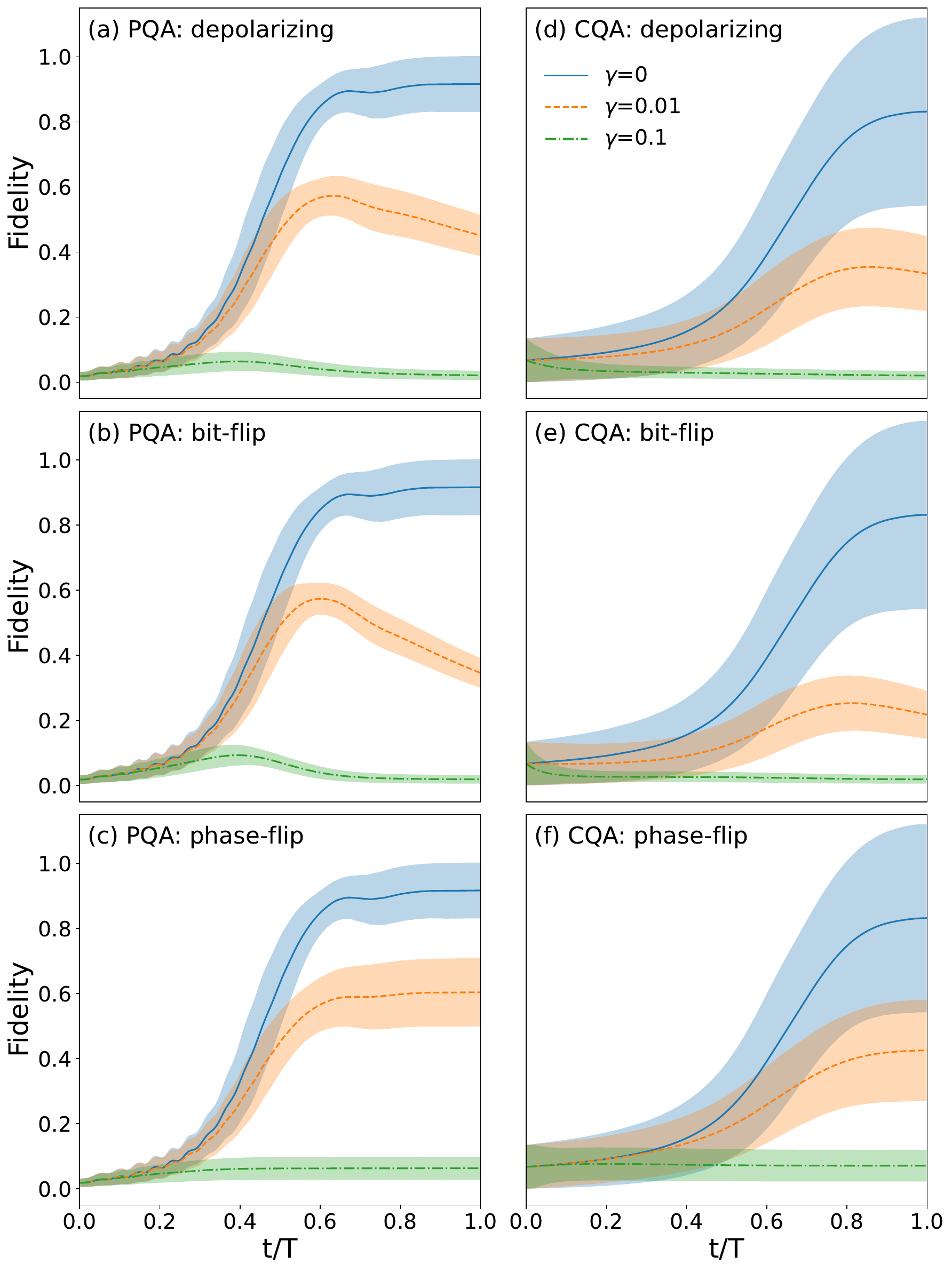}
    \caption{
    Time dependence of the averaged fidelity to the solution state of the 8-node graph partitioning problem. 
    (a)-(c) PQA and (d)-(f) CQA with the depolarising noise (a), (d), bit-flip noise (b), (e), and phase-flip noise (c), (f). 
    We used the number of qubits $N=8$, $\alpha = 2$ and $T= 20$. 
    The average and variance are for 100 samples.
    }
    \label{fig:1}
\end{figure}

Figure~\ref{fig:1} illustrates the time evolution of the averaged fidelity to the optimal solution state for both PQA and CQA under depolarizing, bit-flip, and phase-flip noise. Notably, even in the absence of noise, PQA demonstrates a slight performance advantage over CQA on average. 
Additionally, the variance in fidelity is observed to be broader for CQA compared to PQA, indicating greater fluctuations in the CQA approach. Under depolarizing and bit-flip noise (Figs.~\ref{fig:1}(a)-(b)), an apparent decrease in fidelity is observed in PQA during the latter stages of the annealing process.


\begin{figure}[tbp]
    \centering
    \includegraphics[keepaspectratio, width=80mm]{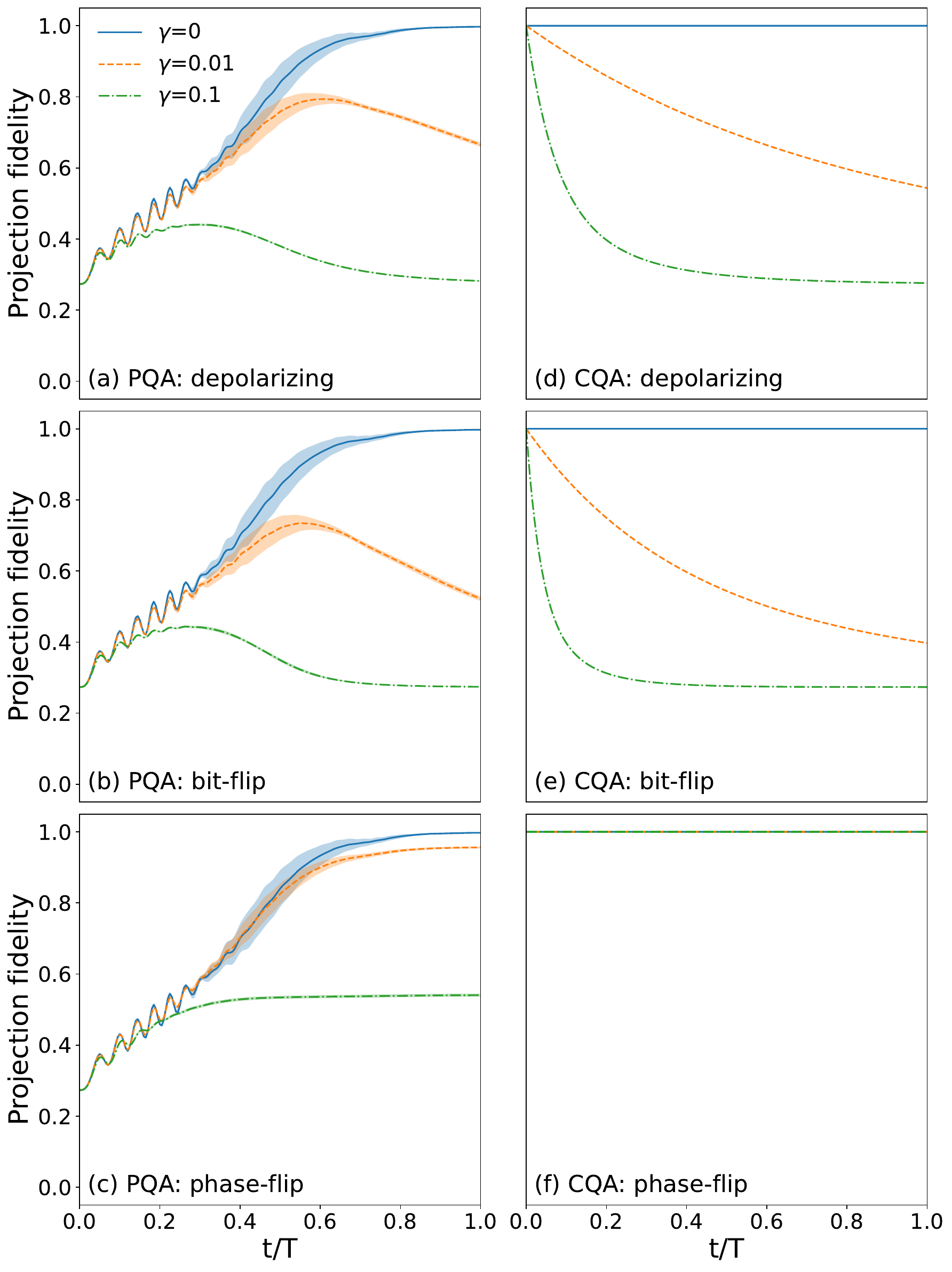}
    \caption{
        Time dependence of the projection fidelity for the subspace ${\mathcal H}_{\rm c}$, given by $ {\rm Tr}[\rho \hat P_{\rm c}]$. Panels (a)-(c) show results for PQA, and panels (d)-(f) for CQA, under depolarizing noise (a), (d), bit-flip noise (b), (e), and phase-flip noise (c), (f). The parameters used are identical to those in Fig.~\ref{fig:1}. 
        The results represent the average and variance over 100 randomly generated instances. 
    }
    \label{fig:2}
\end{figure}


\begin{figure}[tbp]
    \centering
    \includegraphics[keepaspectratio, width=80mm]{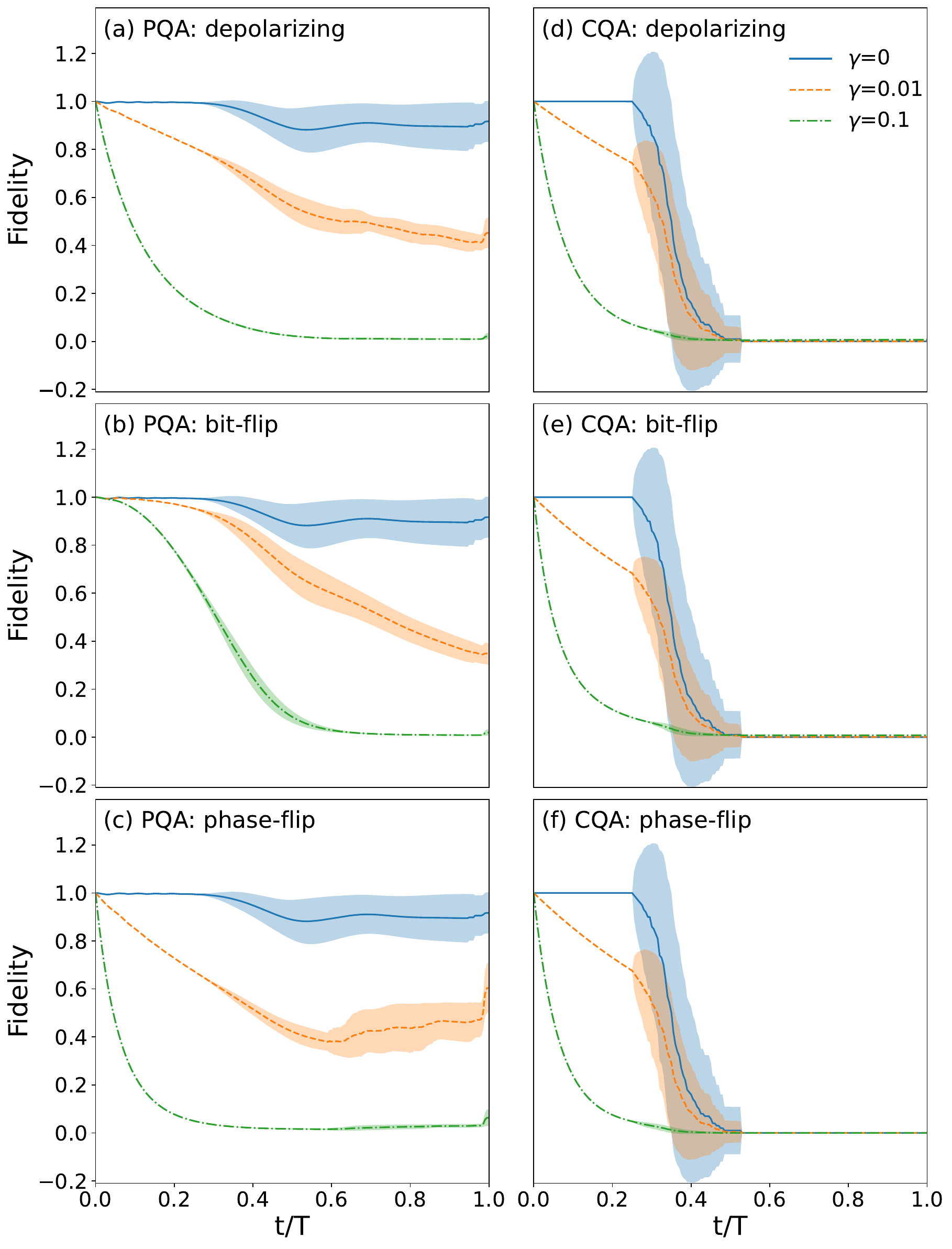}
    \caption{
    Fidelity to the instantaneous ground state of the Hamiltonian. Panels (a)-(c) show results for PQA, and panels (d)-(f) for CQA, under depolarizing noise (a), (d), bit-flip noise (b), (e), and phase-flip noise (c), (f). The parameters used are the same as in Fig.~\ref{fig:1}.
    }
    \label{fig:3}
\end{figure}


\begin{figure}[tbp]
    \centering
    \includegraphics[keepaspectratio, width=60mm]{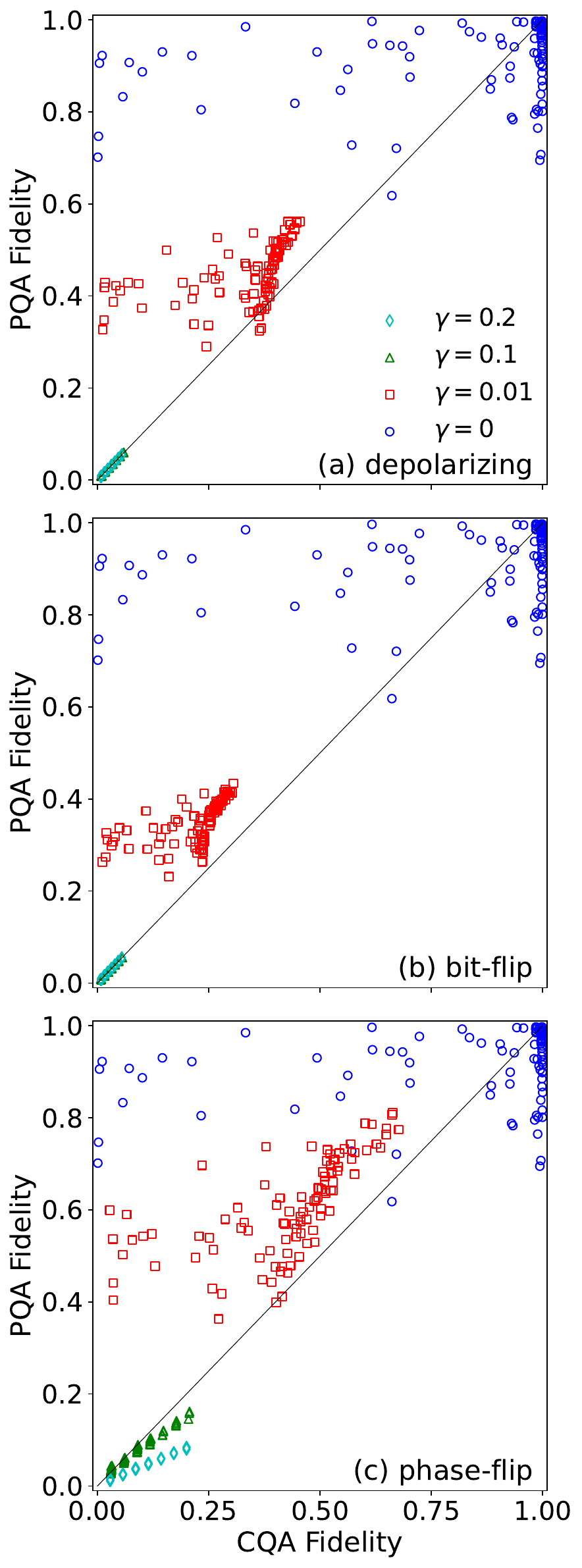}
    \caption{
    Success-probability scatter plots for PQA and CQA. Each point represents the success probabilities of PQA and CQA for the same instance of the graph partitioning problem. Results are shown for (a) depolarizing noise, (b) bit-flip noise, and (c) phase-flip noise. A total of 100 randomly generated instances of the 8-node graph partitioning problem were used in this analysis.
    }
    \label{fig:PQAvsCQA}
\end{figure}

\begin{figure}[tbp]
    \centering
    \includegraphics[keepaspectratio, width=80mm]{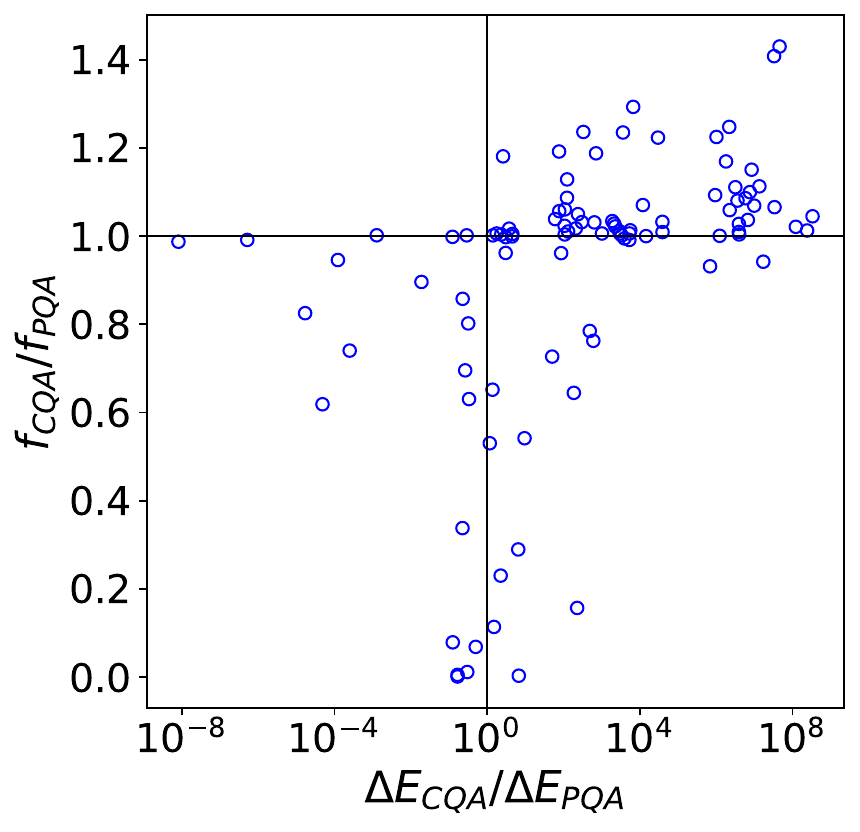}
    \caption{
    Minimum energy-gap ratio versus fidelity ratio between CQA and PQA, where $\Delta E_{\rm CQA(PQA)}$ denotes the minimum energy gap, and $f_{\rm CQA(PQA)}$ represents the fidelity in CQA (PQA), respectively. 
    }
    \label{fig:deltafidleity-deltagap}
\end{figure}


\begin{figure}[tbp]
    \centering
    \includegraphics[keepaspectratio, width=60mm]{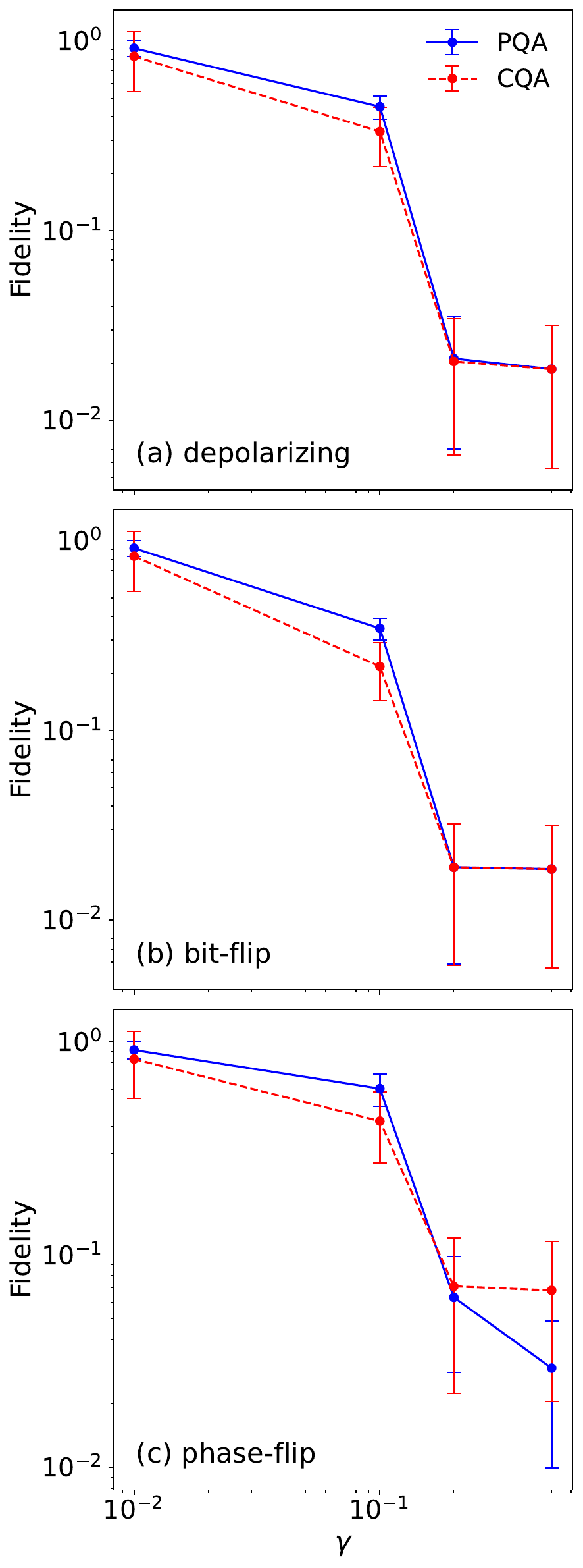}
    \caption{
    Noise strength dependence of the success probability. Results are shown for (a) depolarizing noise, (b) bit-flip noise, and (c) phase-flip noise. The same parameters as in Fig.~\ref{fig:1} were used for these simulations.
    }
    \label{fig:4}
\end{figure}


To quantify the extent of leakage from the constraint-satisfying subspace ${\mathcal H}_{\rm c}$, we calculate the projection fidelity $F_{\rm c} = {\rm Tr} [\rho \hat P_c]$, where $\hat P_c \equiv \sum_{i_{\rm c}} | i_{\rm c} \rangle \langle i_{\rm c}|$ is the projection operator to the subspace ${\mathcal H}_{\rm c}$ the states of which satisfy the constraint, i.e., $C(\{\hat{\sigma}_i^z\}) | i_{\rm c}\rangle = c | i_{\rm c}\rangle $. 
The results are shown in Fig.~\ref{fig:2}. In PQA (Figs.~\ref{fig:2}(a)-(c)), the fidelity $F_{\rm c}$ starts at a lower value due to the exploration of the full Hilbert space $\mathcal H$, but increases over time as the system converges toward the solution state within ${\mathcal H}_{\rm c}$.

In contrast, CQA (Figs.~\ref{fig:2}(d)-(f)) begins with a projection fidelity of $F_{\rm c} = 1$, as the initial state is prepared to fully satisfy the constraint. However, under depolarizing and bit-flip noise, significant leakage occurs from the constrained subspace ${\mathcal H}_{\rm c}$ (Figs.~\ref{fig:2}(d)-(e)), resulting in a marked reduction in the projection fidelity $F_{\rm c}$. 
Notably, phase-flip noise does not induce such leakage, allowing the system to remain within the subspace ${\mathcal H}_{\rm c}$ throughout the entire annealing process (Fig.~\ref{fig:2}(f)). Although the variance across 100 randomly generated instances is shown in Fig.~\ref{fig:2}, it remains small, indicating problem-independence, compared to the fidelity to the solution state shown in Fig.~\ref{fig:1}. 
This occurs essentially because the jump operators are uniformly applied to each spin.
Note that the leakage of symmetry-preserved subspaces has been also discussed in XY quantum alternating operator Ansatz under depolarizing noise~\cite{PhysRevA.103.042412}. 

The absence of the leakage from the subspace ${\mathcal H}_{\rm c}$ in CQA can be understood by considering the fact that 
the commutation relation 
\begin{align}
    [C(\{\hat{\sigma}_i^z\}), \rho] = 0
\end{align}
holds throughout the annealing process if the initial state is a pure state $\rho_0 \equiv |\phi_0\rangle \langle \phi_0|$ that satisfies $C(\{ \hat \sigma_i^z \}) | \phi_0 \rangle = c | \phi_0 \rangle$. 
This implies that the phase-flip noise preserves the constraint, or does not cause the leakage of the density operator $\rho$ from the operator space acting on the subspace ${\mathcal H}_{\rm c}$. 
Indeed, the Lindblad equation updates the density operator $\rho$ over a small time interval $\Delta t$, yielding the following update rule in the phase-flip channel:
\begin{align}
    \rho \to \rho - i \Delta t [\hat{H}_{\rm CQA}, \rho] + \Delta t \sum_i \hat{L}_i \rho \hat{L}_i^\dagger.
\end{align} 
Given that the initial density operator is defined as $\rho_0 = |\phi_0\rangle \langle \phi_0|$, it follows straightforwardly that the updated density operator $\rho$ satisfies the commutation relation $[C(\{\hat{\sigma}_i^z\}), \rho] = 0$. 
Specifically, the noise-free terms commute with the constraint operator, because 
$[C(\{\hat{\sigma}_i^z\}), \rho_0] = 0$ holds as $C(\{\hat{\sigma}_i^z\}) \rho_0 = c \rho_0$, and in CQA, the constraint operator $C(\{\hat{\sigma}_i^z\})$ commutes with both the problem Hamiltonian $\hat{H}_{\rm prob}$ and the driver Hamiltonian $\hat{H}_{\rm d}^{\rm CQA}$, i.e., $[\hat{H}_{\rm CQA}, C(\{\hat{\sigma}_i^z\})] = 0$. 
Furthermore, the phase-flip noise preserves the constraint condition: 
\begin{align}
    [C(\{\hat{\sigma}_i^z\}), \hat{L}_i \rho_0 \hat{L}_i^\dag] = 0,
\end{align} 
since $[C(\{\hat{\sigma}_i^z\}), \hat{L}_i] = 0$ holds for the phase-flip noise $\hat L_i = \sqrt{\gamma} \hat \sigma_i^z$. 

Since the density operator updated from the initial one $\rho_0$ remains in the initial operator space, the subsequent density operator updated by the Lindblad's master equation remains in the same space. 
Therefore, under the influence of phase-flip noise, the evolution of the density matrix is fully contained within the operator space acting on the constraint-satisfying Hilbert subspace ${\mathcal H}_{\rm c}$ throughout the annealing process, ensuring no leakage occurs.
Note that the constraint-satisfying Hilbert subspace ${\mathcal H}_{\rm c}$ differs from the decoherence-free subspace~\cite{PhysRevLett.81.2594}. Although leakage from the constraint subspace is absent under phase-flip noise, states within this subspace remain vulnerable to phase-flip errors, thereby reducing the success probability.


The fidelity to the instantaneous ground state of the Hamiltonian as a function of time is shown in Fig.~\ref{fig:3}. In the PQA (Figs.~\ref{fig:3}(a)-(c)), the fidelity is high during the early stages of annealing ($t/T \lesssim 0.3$), but decreases as the system approaches the minimum energy gap near $t/T \approx 0.5$.  This indicates that the system struggles to follow the instantaneous ground state around the time when the driver Hamiltonian is switched to the problem Hamilotnian. 
The problem-dependence of the fidelity to the instantaneous ground state emerges in the latter stages of the annealing process. 

For CQA (Figs.~\ref{fig:3}(d)-(f)), the fidelity exhibits a sharp decline around $t/T \approx 0.4$. This behavior arises from the fact that the solution state in CQA does not necessarily correspond to the ground state of the problem Hamiltonian $\hat{H}_{\rm prob}$. Specifically, while the ground state of $\hat{H}_{\rm prob}$ satisfies $C(\{\hat{\sigma}_i^z\}) |g\rangle = c_{\rm g} |g\rangle$, it does not always fulfill the required condition $c_{\rm g} = c$. In the graph partitioning case, the ground states of $\hat{H}_{\rm prob}$ in the full Hilbert space are all-spin up or down states, which are orthogonal to target solution states of equal-size partition, or zero $z$-magnetization states. Consequently, the overlap between the solution state and the instantaneous ground state vanishes after the dominant contribution of the Hamiltonian shifts from $\hat{H}_{\rm d}^{\rm CQA}$ to $\hat{H}_{\rm prob}$. The time at which the sharp-drop of the fidelity occurs is problem-dependent, although the absence of overlap in the latter stages of the annealing process is a general feature.


It is widely believed that the enhanced efficiency of the CQA algorithm arises from its comparatively larger energy gap relative to that of PQA~\cite{PhysRevApplied.5.034007}. A wider energy gap between the ground and first-excited states is typically associated with shorter annealing times~\cite{PhysRevApplied.5.034007, BAPST2013127}. However, contrary to our initial expectations, our results reveal that the relative superiority of PQA or CQA is highly problem-dependent (see Fig.~\ref{fig:PQAvsCQA}). 
Under noise conditions, CQA's performance deteriorates substantially, with success probabilities consistently lower than those of PQA, except under strong phase-flip noise conditions.

In the absence of noise, 
the success probabilities of CQA exhibit substantial variability (Fig.~\ref{fig:PQAvsCQA}). Specifically, while CQA successfully identifies the solution state with high probability in some instances, it almost completely fails in others, even under noiseless conditions.
However, the success probabilities of PQA are consistently estimated to exceed about $0.6$, indicating relatively stable performance across different problem instances compared to CQA. 
One of the possible reason is that our expectation of the wider minimum energy-gap in CQA is model-dependent. 
If the wider energy gap remains open in CQA without model-dependence, another plausible explanation for this performance discrepancy lies in the matrix elements that govern the annealing dynamics~\cite{BAPST2013127,PhysRevA.110.022620}. 
In particular, the matrix elements controlling transitions between quantum states in CQA may be larger than those in PQA, potentially inducing detrimental transitions and adversely affecting performance. 


We observe that both scenarios hold (Fig.~\ref{fig:deltafidleity-deltagap}); while many problem instances exhibit a wider energy gap in CQA, as discussed in Ref.~\cite{PhysRevApplied.5.034007}, CQA does not consistently yield a wider minimum-energy gap compared to PQA. Furthermore, even when the minimum energy gap in CQA is larger than in PQA, PQA can still achieve a higher success probability. This observation may support the scenario where matrix elements, often not considered in performance discussions, play a significant role~\cite{PhysRevA.110.022620}.


The dependence of the success probability on noise strength is summarized in Fig.~\ref{fig:4}. As expected, the success probability decreases monotonically with increasing noise strength across all noise models. 
In the weak noise regime, PQA outperforms CQA in terms of success probability on average. 
However, for relatively strong phase-flip noise, CQA achieves higher success probabilities than PQA (Fig.~\ref{fig:4}(c)).
It is important to note that while CQA performs better in the presence of strong phase-flip noise, the overall success probability remains low. 
This suggests that although CQA offers certain advantages in phase-flip noise, it is not yet a practical solution for noisy quantum annealers, especially under realistic noise conditions where control over the noise model is limited.

CQA offers potential advantages in hardware efficiency, as it eliminates the need for the all-to-all connectivity required by penalty terms in PQA. 
While in certain problem instances, of course, CQA can indeed outperform PQA, our findings indicate that PQA demonstrates stability in terms of success probability, both in the absence and presence of noise. 
For CQA, leakage from the constraint-satisfying subspace is inevitable across most noise models, with the exception of phase-flip noise. 
To mitigate leakage during the annealing process, approaches such as error-corrected quantum annealing~\cite{PhysRevLett.100.160506,Pudenz2014} and autonomous error correction~\cite{PhysRevA.72.012306,PhysRevLett.111.120501,Mirrahimi_2014,PhysRevA.90.062344,PhysRevLett.116.150501,PhysRevA.98.012317,Albert_2019,Ma2020,Gertler2021,PRXQuantum.3.020302,Kwon2022} are essential. In particular, in order to confine the system within the constraint subspace, suppressing bit-flip noise is imperative. The cat code~\cite{CAI202150,Lescanne2020,Iyama2024,hoshi2024entanglingschrodingerscatstates,PRXQuantum.3.010329} is promising, as it exponentially suppresses bit-flip errors while the phase-flip error scales linearly with the cat size~\cite{Lescanne2020}. 
These developments are key to unlocking the full potential of CQA in noisy quantum devices.

\section{Conclusion}
We systematically evaluated the performance of penalty-based quantum annealing (PQA) and constrained quantum annealing (CQA) in solving the graph partitioning problem under various noise models, including depolarizing, bit-flip, and phase-flip noise. 
Contrary to our expectations, 
CQA does not always outperform PQA even in the absence of the noise. 
Our results underscore the problem-dependent nature of quantum annealing. 
In the absence of noise, PQA demonstrates relatively stable success probabilities across problem instances, whereas CQA exhibits significant variability. 

Under noisy conditions, PQA consistently outperforms CQA, especially with depolarizing and bit-flip noise. 
Furthermore, CQA suffers from substantial leakage from the constraint-satisfying subspace, which degrades its performance. Interestingly, phase-flip noise does not induce such leakage in CQA, resulting in better performance under strong phase-flip noise. However, in this regime, the overall success probabilities remain low, limiting the practicality of CQA for current noisy quantum devices.

Although CQA is hardware efficient, PQA shows greater resilience to noise and more consistent performance. To fully exploit the potential of CQA, future work must focus on developing error-corrected quantum annealing and autonomous error correction to mitigate leakage from the constraint-satisfying subspace during the annealing process.
The cat code is promising for CQA as it exponentially suppresses bit-flip errors.

\section*{Acknowledgement}
S.W. was supported by JST, PRESTO Grant Number JPMJPR211A, Japan.
Y.M. was supported by JST, PRESTO Grant Number JPMJPR2018, Japan.
The authors thank T. Kanao for useful discussion. 

\bibliography{ref}

\end{document}